\documentclass[twoside]{article}
\usepackage{graphicx}
\usepackage{fancyhdr}
\usepackage{multicol}
\usepackage{styl-ap}

\begin{document}
\title{Star Formation in Tadpole Galaxies}
\author{C.~Mu\~noz-Tu\~n\'on\work{1,2}, J.~S\'anchez~Almeida\work{1,2}, D.~M.~Elmegreen\work{3} \& B.~G.~Elmegreen\work{4}}
\workplace{Instituto de Astrof\'\i sica de Canarias, E-38205 La Laguna,
Tenerife, Spain
\next
Departamento de Astrof\'\i sica, Universidad de La Laguna,
Tenerife, Spain
\next
Department of Physics and Astronomy, Vassar College,
Poughkeepsie, NY 12604, USA
\next
IBM Research Division, T.J. Watson Research Center, Yorktown Heights,
NY 10598, USA}
\mainauthor{cmt@iac.es}
\maketitle

\begin{abstract}%

Tadpole Galaxies look like a star forming head with a tail structure to the side.
They are also named cometaries. In a series of recent works we have discovered a
number of issues that lead us to consider them extremely interesting targets.
First, from images, they are disks with a lopsided starburst. This result is firmly
established with long slit spectroscopy in a nearby representative sample. They
rotate with the head following the rotation pattern but displaced from the
rotation center. Moreover, in a search for extremely metal poor (XMP) galaxies, we
identified tadpoles as the dominant shapes in the sample -- nearly 80\% of the
local XMP galaxies have a tadpole morphology. In addition, the spatially resolved
analysis of the metallicity shows the remarkable result that there is a
metallicity drop right at the position of the head. This is contrary to what
intuition would say and difficult to explain if star formation has happened from
gas processed in the disk. The result could however be understood if the star
formation is driven by pristine gas falling into the galaxy disk. If confirmed, we
could be unveiling, for the first time, cool flows in action in our nearby world.
The tadpole class is relatively frequent at high redshift -- 10\% of resolvable
galaxies in the Hubble UDF but less than 1\% in the local Universe. They are
systems that could track cool flows and test models of galaxy formation.
\end{abstract}


\begin{multicols}{2}
\section{Introduction}

Elongated galaxies with bright clumps at one end are visible in deep
field images taken with HST or from the ground; van der Bergh et al., 1996 called
them ``tadpole'' galaxies". Figure 1 in Elmegreen et al. (2005) shows different morphologies of galaxies observed with
the Hubble Ultra Deep Field (UDF). The fourth row presents images of Tadpoles.
This asymmetric morphology is rather common at high redshift but rare in the local
universe. For example, tadpoles constitute 10\% of all galaxies larger than 10
pixels in the UDF (Elmegreen et al. 2007; Elmegreen \& Elmegreen, 2010), and they
represent 6\% of the UDF galaxies identified by Straughn et al. (2006) and
Windshorst et al. (2006) using automated search algorithms.

In contrast, Elmegreen et al. (2012; hereafter Paper I), find only 0.2\%\ tadpoles
among the uv-bright local galaxies of the Kiso survey by Miyauchi-Isobe et al.
(2010). This decrease suggests that the tadpole morphology represents a common but
transition phase during the assembly of some galaxies. Since local tadpole
galaxies are very low mass objects compared to their high redshift analogues, this
phase must be already over for the local descendants of high redshift tadpoles.

The tadpole structure has inspired several explanations, such as ram pressure
stripping that triggers star formation at the leading edge, to mergers (see
S\'anchez~Almeida et al. 2013 for an extensive review). The explanation that we
propose, based on observational evidence summarized here, is that the starburst
head may result from the accretion of an external flow of pristine gas that
penetrates the dark matter halo and hits and heats a pre-existing disk, which is
viewed to the side as the tail.

We will briefly present a summary of recent results showing, first that the local
Tadpoles share the properties of their higher redshift and higher mass
counterparts (section 2), that they belong to the extremely metal poor (XMP)
sample of the Blue Compact Dwarfs family (section 3), they are rotating discs
(section 4) and, finally, that the starbursts (heads) show a drop in the already
low metallicity that can only be understood if fresh metal-poor gas is falling
onto the galaxy (section 5). We finish with a brief summary and future actions.

\section{Local and hight z TPG share properties.}

We used Sloan Digital Sky Survey data to determine the ages, masses, and surface
densities of the heads and tails in 14 local tadpoles (shown in
Figure~\ref{CMT-fig1}) selected from the Kiso and Michigan surveys of UV-bright
galaxies, and we compared them to Tadpoles previously studied in the Hubble Ultra
Deep Field.  The result, published in Paper~I is that the young stellar mass in
the head scales linearly with the rest-frame galaxy luminosity, ranging from
$\approx 10^5$ $M_\odot$ at galaxy absolute magnitude $U = -13$ mag to $10^9$
$M_\odot$ at $U = -20$ mag. The tails in the local sample look like bulge-free
galaxy disks. Their photometric ages decrease from several Gyr to several hundred
Myr with increasing redshift. The far-outer intensity profiles in the local sample
are symmetric and exponential. We suggest that most local tadpoles are bulge-free
galaxy disks with lopsided star formation, perhaps from environmental effects such
as ram pressure or disk impacts, or from in-situ gas collapse to a giant
star-forming clump with a Jeans-length that is comparable to half the disk size.
The existence of a disk, proposed from the analysis of the luminosity profiles,
has been further confirmed spectroscopically (see section 4).

\begin{myfigure}
\centerline{\resizebox{70mm}{!}{\includegraphics{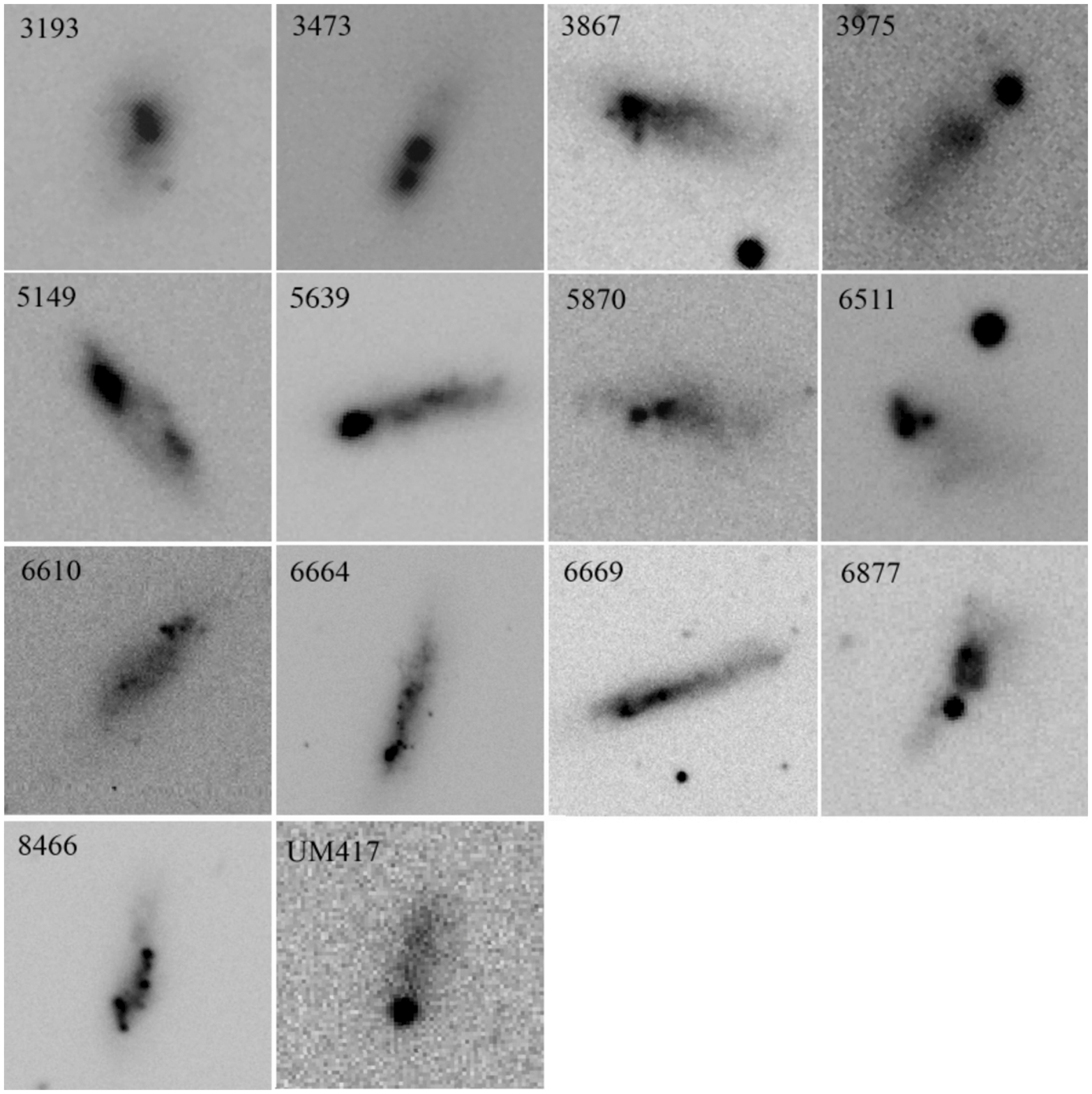}}}
\caption{Tadpole galaxies from the Kiso and UM samples. Figure from Elmegreen et
al., 2012.} \label{CMT-fig1}
\end{myfigure}

\section{Belong to the BCDs- XMP class.}

Blue Compact Galaxies are important targets for a number of reasons. They are
small systems, thought to remain as fossils or debris from the formation of larger
galaxies in the early epochs of the Universe. Their low metallicities, high
sSFRs and low doubling time (less than 1Gy) and active stabursts make them ideal targets to
study the old Universe at low redshift. Their complete census, as well as the
likelihood that they have a long quiescent phase between starbursts, took us to
study the whole sample by making use of SDSS. The properties of the quiescent and
burst phases were determined from detailed studies of the bursts and host galaxies
of a nearby sample (Amorin et. al., 2007, 2008). These properties allowed us to
identify such objects in the database. The SDSS DR6 database provides $\approx$
21.500 quiescent BCD candidates, a number 30 times larger to those bursting
(BCDs). This result implies that one out of every three dwarf galaxies in the
local universe may be a quiescent BCD. The properties of the two samples are
consistent with a single sequence in galactic evolution with the quiescent phase
lasting 30 times longer than the burst phase (S\'anchez~Almeida et. al., 2008). In
S\'anchez~Almeida et al. (their figure 9) there is a clear subsample of objects
with the lowest luminosity which turn out to be also those that are most metal
poor.

\begin{figure*}
\includegraphics[width=\textwidth]{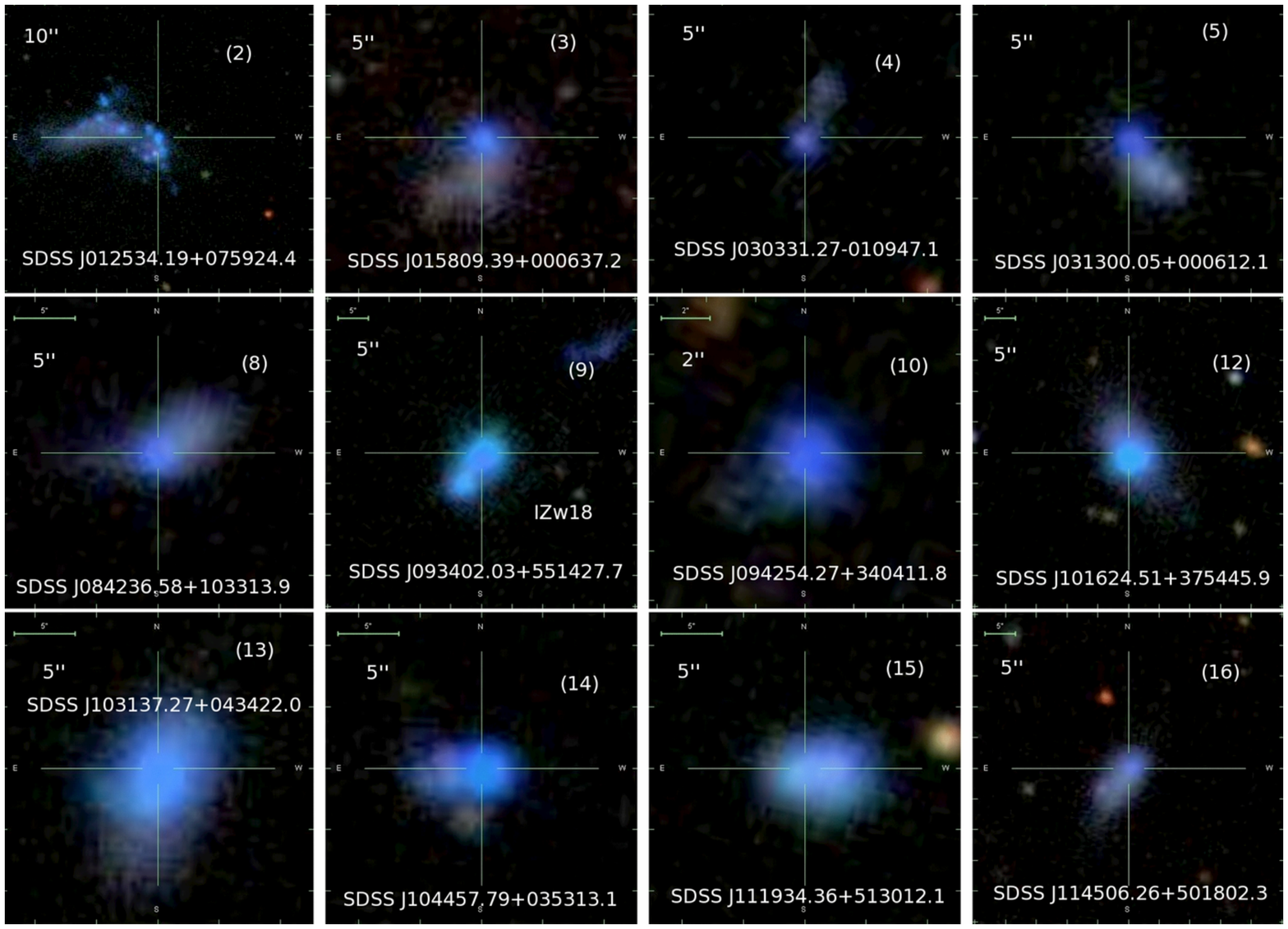}
\caption{SDSS mugshots of all XMP candidates (from spectroscopy). The figure is
taken from Morales-Luis et al., 2011. Note the prevalence of Tadpole shapes.}
\label{CMT-fig2}
\end{figure*}
%%%%%%%%%%%%%%%%%

We carried out a systematic search for extremely metal-poor (XMP) galaxies in the
spectroscopic sample of Sloan Digital Sky Survey (SDSS) data release 7 (DR7) (Morales-Luis et al., 2011). The
XMP candidates are found by classifying all of the galaxies according to the form
of their spectra in a region 80 $\AA$ wide around H$\alpha$ using an automatic
classification algorithm, {\it k-means} (S\'anchez~Almeida et al.,2009). Our systematic search renders 32 galaxies
having negligible [N II] lines, as expected in XMP galaxy spectra. Twenty-one of
them were previously identified as XMP galaxies in the literature, and 11 were
new. This was established after a thorough bibliographic search that yielded only
some 130 galaxies known to have an oxygen metallicity 10 times smaller than the
Sun (explicitly, with 12 + log [O/H] $\leq$ 7.65). XMP galaxies are rare; they
represent 0.01\% of the emission lines galaxies in SDSS/DR7. The XMP galaxies
constitute 0.1\% of the galaxies in the local volume, or $\approx$ 0.2\% of the
emission-line galaxies. All but four of our candidates are BCD galaxies, and 24 of
them have either cometary shape or are formed by chained knots. Note that this
result, the morphology, is absolutely independent of our search criterium
(spectra).

%\begin{figure}[htb]
%\centerline{\includegraphics{figure.eps}}
%\caption{Big figure}
%\label{author-fig2}
%\end{figure}

\section{Tadpoles are Rotating disks}

The work in Paper I was followed up by mean of high spectral resolution long slit
observations at the INT (2.5m) at Observatorio del Roque de los Muchachos (ORM)
with IDS spectrograph. In order to determine the dynamical properties and
metallicities of local tadpoles, we measured H$\alpha$ spectra along the head-tail
direction in a representative fraction of the original  sample. Further details
are in S\'anchez~Almeida et al. (2013). In Figure~\ref{CMT-fig3}, we show spectral
flux (the solid lines) and H$\alpha$ flux (the dashed lines) along the slit. Note
the obvious lopsidedness of the light distributions, as expected from the tadpole
shape. The origin of distance on the abscissa has been set as the position of the
maximum of the spectral flux distribution. The dotted line represents a Gaussian
fitted to the H$\alpha$ flux around the head. The horizontal bar in each panel
gives a common length scale corresponding to 1 kpc. The thin vertical solid lines
indicate the center of rotation obtained from the rotation curve fit shown in
Fig.~\ref{CMT-fig4}.

Velocities,  masses, abundances and other physical parameters were determined from
the spectra. Bulk velocities were measured from the displacement of H$\alpha$. We
computed the displacement both as the barycenter of the emission line, and as the
center of a Gaussian function fitted to the profile. Errors were estimated from
the S/N measured in the continuum and then propagated to the centroids. The FWHM
of the profiles were also measured directly from the profiles and from the
Gaussian fits. Their errors were also inferred from the noise measured in the
continuum by error propagation.

Figure~\ref{CMT-fig4} shows the velocity curves of the tadpole galaxies included
in the study. The abscissae represent distances along the major axes of the
galaxies from the position of the tadpole head; i.e., the brightest point on the
galaxy. The range of distances differs for the different targets, but the
horizontal bar in each panel gives a common length scale corresponding to 1 kpc.
The points with error bars show the observations whereas the thick solid line
represents the best fit of the observed points to the analytic rotation curve. The
part of the rotation curve shown in red indicates the portion of the velocity
curve used for fitting. The thin horizontal and vertical lines indicate the
systemic velocity and the center of rotation obtained from the fit, respectively.
The little arrows on top of each panel also indicate the centers of rotation. The
zero of the velocity scale is set by velocity of the spatially integrated spectrum
of the galaxy; positive velocities are redshifts.

The main result shown in Figure~\ref{CMT-fig4} is that five out of seven targets
show velocity gradients interpreted as rotation.  Another important result is that
the tadpole head is not at the rotation center.

Sometimes the interpretation of the velocity curve as rotation is obvious (e.g.,
{\sc kiso8466}), but other times the curve looks more like a perturbed rotation
(e.g., {\sc kiso5639}).  {\sc kiso3193} and {\sc kiso3867} have a rather flat
velocity curve, and therefore no obvious rotation. However, one of them, {\sc
kiso3867}, shows a systematic line shift of the order 10--20~km\,s$^{-1}$ between
the two extremes of the galaxy (Figure~\ref{CMT-fig4}). The amplitude is of the
order of the error bars, but the displacement is in the raw data as judged by
inspection of the individual H$\alpha$ profiles.

%%%%%%%%%%%%%%%%%%%%%%%%%%%%%%%%%%%%%

\begin{myfigure}
\centerline{\resizebox{70mm}{!}{\includegraphics{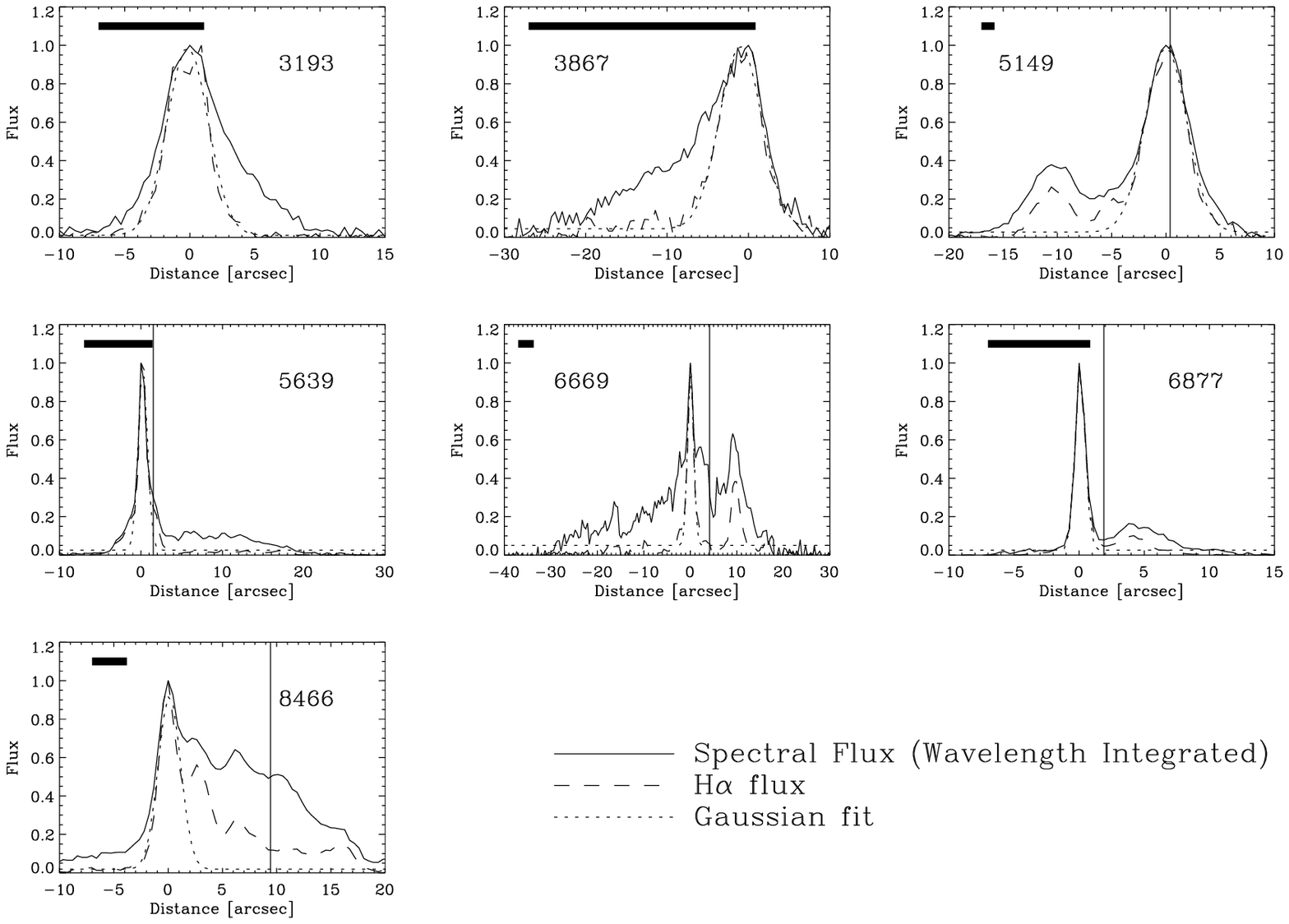}}}
\caption{Spectral flux (the solid lines) and H$\alpha$ flux (the dashed lines).
Formal error bars for photometry are not included since they are negligibly
small.} \label{CMT-fig3}
\end{myfigure}

%%%%%%%%%%%%%%%%%%%%%%%%%%%%%%%%%%%%%%

\begin{myfigure}
\centerline{\resizebox{70mm}{!}{\includegraphics{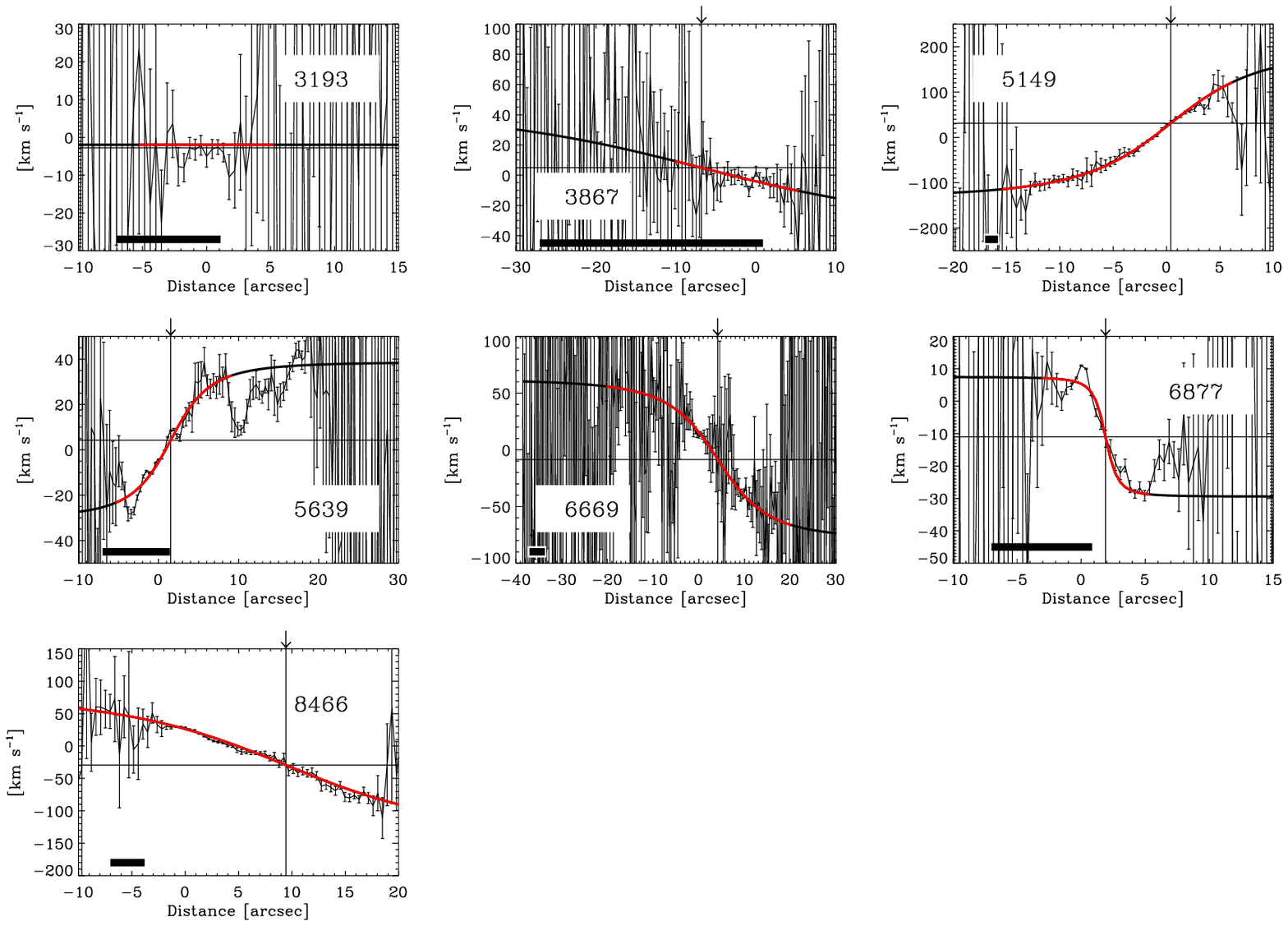}}}
\caption{Velocity curves of the seven tadpole galaxies obtained from long slit
spectroscopy. Figure from S\'anchez~Almeida et al. 2013} \label{CMT-fig4}
\end{myfigure}

\section{Drops in metallicity at the head location}

\begin{figure*}
\includegraphics[width=\textwidth]{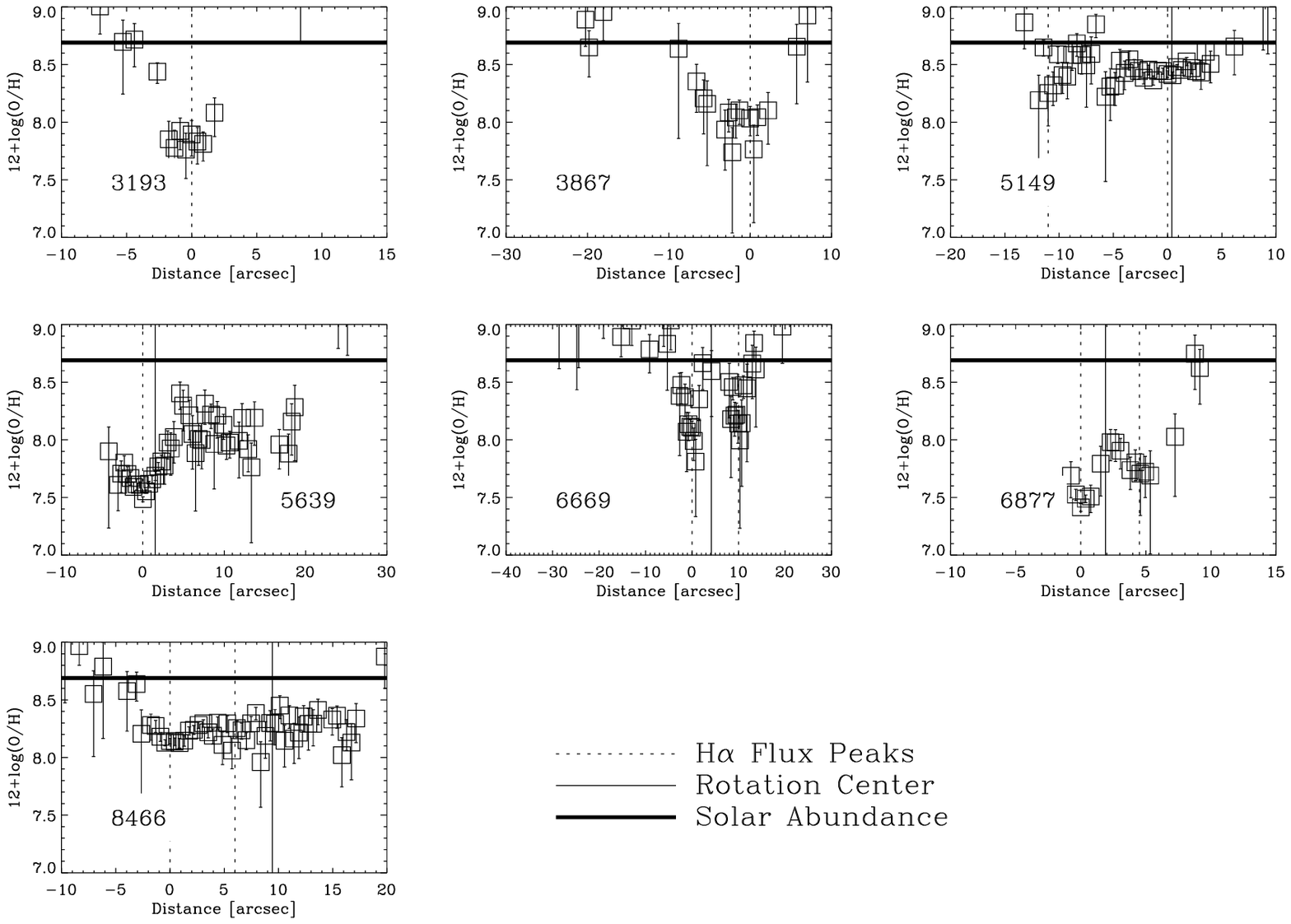}
\caption{Oxygen abundance variation across the galaxies. The vertical solid line
represents the center of rotation, whereas the vertical dotted line indicates the
location of the maximum in the H$\alpha$ flux profile. The thick horizontal solid
line indicates the solar metallicity.
Note the existence of abundance variations, with the  minima coinciding with the
largest H$\alpha$ signals. Note also that {\sc kiso5639} and {\sc kiso6877} reach
very low abundances, below one-tenth of the solar value; therefore, they are
members of the set of rare  XMP galaxies. } \label{CMT-fig6}
\end{figure*}

The metallicity along the slit was estimated using the ratio [NII]$\lambda$6583 to
H$\alpha$. Figure~\ref{CMT-fig6} taken from S\'anchez~Almeida et al (2013) shows
the variation across the galaxies of the oxygen abundance, including their error
bars. All galaxies have sub-solar metallicity - the thick horizontal line marks
the solar oxygen abundance given by $12+\log({\rm O/H})_\odot=8.69\pm 0.05$
(Asplund et al., 2009). The galaxies also present significant abundance gradients,
with the lowest abundances tending to coincide with the largest H$\alpha$
emissions (e.g., {\sc kiso6669} and {\sc kiso6877}) in Fig.~\ref{CMT-fig6}.
Remember that the vertical dotted lines mark the position of the peak H$\alpha$
fluxes. Two targets, {\sc kiso5639} and {\sc kiso6877}, have metallicities well
below one-tenth the solar value, therefore, they belong to the XMP galaxies class
(Kunth \& \"Ostlin, 2000).

XMP galaxies are really rare objects: one out of a thousand galaxies in the local
universe according to Morales-Luis et al. (2011). Therefore the fact that we
observe two in a sample of seven cannot be a coincidence. It is known that a
significant fraction of XMP galaxies turn out to be cometary or tadpole (Papaderos
et al., 2008). We have found that the reverse holds too, i.e., that tadpole
galaxies have a significant probability of being XMP. This fact supports the idea
that the tadpole morphology is a sign of dynamical youth, as the low metallicity
is a sign of being chemically young (see discussion in S\'anchez~Almeida et al.,
2013).

%%%%%%%%%%%%%%%%%

%%%%%%%%%%%%%%%%%

\section{Sumary and future}

Tadpole Galaxies have an easily identified shape, belong to the Blue Compact Dwarf
class, and are rare in the nearby Universe. Their appearance suggests then to be
the consequence of some interaction or star formation triggered by ram pressure
processes. These local tadpoles seem though to form a continuous sequence with the
UDF Tadpoles, seen in relatively larger number at high redshift (Elmegreen \&
Elmegreen 2010). Regarding their photometric properties, local Tadpoles occupy the
low mass end in sequences such as star formation, surface density and
mass-to-light ratio. In addition, the radial intensity profiles of both samples
(local and high redshift Tadpoles) show an exponential decrease at large
galactocentric distances, which we interpreted as an evidence for the existence of
an underlying disk.

From the point of view of their chemical content, Extremely Metal Poor (XMP)
galaxies  are the least evolved objects in the local universe (Pagel et al., 1992;
Kunth \& \"Ostlin, 2000). They represent only 0.1 \% of the galaxies in an
arbitrary nearby volume (Morales-Luis et al., 2011) and a significant fraction of
these chemically primitive objects turn out to have tadpole or cometary shape
(Brinchmann et al. 2008). This association between low metallicity and tadpole
shape suggests that both are attributes of very young systems.

Detailed studies of a significative sample of the local Tadpole class by means of
high-resolution long-slit spectroscopy put into evidence two important facts: they
do rotate and they show oxygen abundances that vary along the disk.  A high
percentage of them are XMP galaxies and the metallicity is minimum were
star-formation is maximum. Moreover, the rotation center does not coincide with
the current starburst location (the head).

The  oxygen metallicity estimated from [NII]6583/H$\alpha$ often shows significant
spatial gradients across the galaxies ($\sim$0.5~dex), being lowest at the head
and increasing in the rest of the galaxy, tail included. A similar result with the highest SFR 
been the lowest metallicity region has also been found in low metallicity GRB host galaxies
 (Levesque et al., 2011).

The sense of the resulting
metallicity gradient differs from the observation of local disk galaxies, where
the gas-phase metallicity increases toward the galaxy centers (Vilchez et al.,
1988) or is just constant (Genel et al., 2008). However, the type of variation we
measure, with a minimum metallicity at the most intense star-forming region, has
been observed in galaxies at redshift around 3 by Cresci et al. (2010), where it
is interpreted as evidence for infall of pristine gas triggering star formation.

As a result of all the evidence found, we propose that local Tadpole Galaxies are
disks still in the process of being formed in our present-day Universe. Moreover
their otherwise very low star formation is enhanced and triggered by cool-flow
accretion of pristine metal-poor gas. This triggering causes an extremely
metal-poor head to shows up on the side of an immature (still forming) rotating
disk.

There are a number of parallel actions now being undertaken to go further.
We have just finished a study of the HI content of a complete sample of XMP
galaxies (Fihlo et al., 2013) and compared the properties of the clumps of different galaxy types at different
redshifts (Elmegreen et al., 2013). The recent analysis of a new sample corroborates the 
metallicity drop in the XMP galaxies using the so call direct method (S\'anchez~Almeida et al. 2014).

Cosmological simulations predict cold-flow buildup to be the main mode of galaxy
formation (Dekel et al., 2009). The incoming gas is expected to form giant clumps
that spiral in and merge into a central spheroid. To confirm such a scenario at
high redshift is neither easy nor feasible and we think we have discovered nearby
cases that may allow for comparisons.

We are making a comprehensive study of the
disks and bursts of other BCDs with XMP properties to search for extremely metal
poor clumps in 2D spectroscopy. The final aim is to characterize the local XMP
Galaxy sample within the cool flow paradigm.

\thanks
This work has been partly funded by the Spanish Ministry for Science, project
AYA~2010-21887-C04-04. Results based on observations at the Observatorio del Roque
de los Muchachos (ORM), operated by the IAC at La Palma. Thanks to Franco
Giovannelli and Lola Sabau-Graziatti, organizers of this workshop for their
kindness and dedication. Thanks to the referee for useful comments.

\end{multicols}
\end{document}